\documentclass[twocolumn,aps,prl,superscriptaddress
,secnumarabic
]{revtex4}
\usepackage{bm,amsmath,amsfonts,amssymb,esint,graphicx,color,MnSymbol}

\newcommand{\Eq}[1]{Eq.~\eqref{#1}}
\newcommand{\eq}[1]{\eqref{#1}}
\newcommand{\Fig}[1]{Fig.~\ref{#1}}

\newcommand{\beq}{\begin{equation}}
\newcommand{\eeq}{\end{equation}}
\newcommand{\beqa}{\begin{eqnarray}}
\newcommand{\eeqa}{\end{eqnarray}}
\newcommand{\Beqa}{\begin{eqnarray*}}
\newcommand{\Eeqa}{\end{eqnarray*}}
\newcommand{\nn}{\nonumber}
\newcommand{\ds}{\displaystyle}

\newcommand{\pdag}{{\phantom{\dagger}}}
\newcommand{\pprime}{{\phantom{\prime}}}
\newcommand{\past}{{\phantom{\ast}}}
\newcommand{\msk}{\mkern 2mu}
\newcommand{\nmsk}{\mkern -2mu}
\newcommand{\smsk}{\mkern 1mu}
\newcommand{\snmsk}{\mkern -1mu}

\begin{document}

\title{Quantum Criticality in Resonant Andreev Conduction}
\author{M. Pustilnik}
\affiliation{School of Physics, Georgia Institute of Technology, Atlanta, Georgia 30332, USA}
\author{B. van Heck}
\affiliation{Department of Physics, Yale University, New Haven, Connecticut 06520, USA}
\author{R. M. Lutchyn}
\affiliation{Station Q, Microsoft Research, Santa Barbara, California 93106-6105, USA}
\author{L. I. Glazman}
\affiliation{Department of Physics, Yale University, New Haven, Connecticut 06520, USA}

\begin{abstract}
Motivated by recent experiments with proximitized nanowires, we study a mesoscopic $s$-wave superconductor connected via point contacts to normal-state leads. We demonstrate that at energies below the charging energy the system is described by the two-channel Kondo model, which can be brought to the quantum critical regime by varying the gate potential and conductances of the contacts.
\end{abstract}

\maketitle
The prediction of and search for Majorana physics in hybrid semiconductor-superconductor structures~\cite{Majorana_reviews} touched off a rapid progress in the technology of such devices \cite{Mourik2012,Rokhinson2012,Das2012,Deng2012,Plissard2012,Plissard2013,Finck2013,Churchill2013,Krogstrup2015,Higginbotham2015,Albrecht2016,Deng2016,Zhang2016,Chen2016,Gul2017,Chang2015,Taupin2016,Sherman2017}. In particular, the pairing gap induced in semiconductor wires by the proximity effect is already comparable with that in a bulk superconductor. 

When a proximitized wire with spin-orbit coupling is placed in a sufficiently strong magnetic field, the nature of the induced  superconducting pairing changes from \mbox{$s$ wave} to $p$ wave~\cite{RL,FvO}, leading to the appearance of Majorana zero modes~\cite{RG2000}. These modes make possible resonant electron transport through a proximitized segment contacted by normal-state leads~\cite{Fu2010,HLG,LG}. Both the width and the height of the resonant Coulomb blockade peaks in the dependence of the conductance on the gate potential saturate at low temperature~\cite{HLG,LG,footnote}. The height of the peaks in this limit is controlled by the asymmetry between the contacts, reaching $e^2\nmsk/h$ in a symmetric device~\cite{HLG,LG}.  For this behavior to be viewed as a signature of the presence of Majorana modes, it must differ from that in the regimes when the Coulomb-blockaded segment is either in the normal state or in the conventional $s$-wave superconducting state. 

In this paper we show that the behavior of the conductance in the $s$-wave regime is not only very different from that described above, but is interesting in its own right. Indeed, it turns out that tunable proximitized devices are ideally suited for the observation of the two-channel Kondo effect, with two almost degenerate charge states of the proximitized segment playing the part of the two states of spin-1/2 impurity. The shape of the Coulomb blockade peaks depends strongly on the asymmetry between the contacts. In a fine-tuned symmetric device the width of the peaks scales at low temperature as~$\sqrt{T}$, whereas their height approaches $2e^2\nmsk/h$. This behavior is a manifestation of the quantum criticality inherent in the two-channel Kondo model. On the contrary, in a generic device with asymmetric contacts, the conductance is proportional to $T^2$ for any gate potential, and vanishes at zero temperature.

We model the system by the Hamiltonian  
\beq
H = H_0 + H_S + H_C + H_T.
\label{1}
\eeq
The first term here describes electrons in the leads. It reads 
\mbox{$H_0= \sum_{\alpha k \sigma}^\pdag \xi^\pdag_k c^\dagger_{\alpha k \sigma}c^\pdag_{\alpha k \sigma}$}, where \mbox{$\alpha = R,L$} labels the right/left lead and $\sigma = \uparrow,\downarrow$  labels the spin. (We will also use the notation $\sigma = \pm 1$ for the spin indices.) In order to study transport at low temperatures, it is adequate to linearize the single-particle spectra as $\xi_k = vk$. Here $v$ is the Fermi velocity and $k$ are the momenta measured from the respective Fermi levels. (We work in units where \mbox{$\hbar = 1$.}) The second term in \Eq{1}, $H_S$, describes an isolated superconductor. In the conventional BCS framework, it is given by~\cite{Tinkham}
\beq
H_S = \sum_{n\sigma}\sqrt{\Delta^2 + \varepsilon_n^2\msk}\smsk
\gamma^\dagger_{n\sigma}\gamma^\pdag_{n\sigma},
\label{2}
\eeq   
where $\Delta$ is the superconducting gap, $\gamma_{n\sigma}$ is the fermionic quasiparticle operator and $\varepsilon_n$ are single-particle energies characterized by the mean level spacing $\delta\ll \Delta$. The third term in \Eq{1} originates in electrostatics and is given by
\beq
H_C = E_C (\hat N - N_{\snmsk g})^2,
\label{3}
\eeq
where $E_C\ll \Delta$ is the charging energy, $N_{\snmsk g}$ is the dimensionless gate potential, and $\hat N$ is an operator with integer eigenvalues representing the number of electrons in the superconductor. Finally, $H_T$ describes the tunneling,
\beq
H_T = \!\!\!\sum_{N\nmsk\alpha n\sigma}\!\!t^\past_{\alpha n^\past}\!\! \nmsk c^\dagger_{\alpha\sigma}\snmsk(0)\smsk d^\past_{n\sigma^\past}\!\!\nmsk
|N-1\rangle\langle N| + \text{H.c.}
\label{4}
\eeq
Here $t_{\alpha n}$ is the tunneling amplitude, the operator
$c^\dagger_{\alpha\sigma}\snmsk(0) = L^{-1/2}\snmsk \sum_k^\pdag\nmsk c^\dagger_{\alpha k\sigma}$ creates an electron with spin~$\sigma$ at point contact $\alpha$ ($L$ is the size of the system that will be taken to infinity in the thermodynamic limit), \mbox{$d_{n\sigma} = u_n\gamma_{n\sigma} -\sigma v_n\gamma^\dagger_{n,-\sigma}$}, where the BCS coherence factors $u_n$ and $v_n$ satisfy~\cite{Tinkham}
\mbox{$u_n^2 = 1- v_n^2 = \frac{1}{2}\bigl(1 - \varepsilon_n/\nmsk\sqrt{\Delta^2 + \varepsilon_n^2}\msk\bigr)$},
and $|N\rangle$ is eigenvector of $\hat N$ with eigenvalue $N$.
 
At low temperatures $T\ll\Delta$, the superconductor favors states with an even number of electrons $N$. Taking into account virtual transitions to states with odd~$N$~\cite{Tinkham,HGMS} in the second order of perturbation theory, we obtain
\mbox{$H = H_0 + H_C + H_A$}, where
\beq
H_{\snmsk A} = \!\sum_{\text{odd}\, N}\sum_{\,\alpha\alpha'}
J_{\alpha\alpha'}\smsk c_{\alpha\uparrow}(0)\smsk c_{\alpha'\snmsk\downarrow}(0)
\msk|N+1\rangle\langle N - 1| + \text{H.c.}
\label{5}
\eeq
describes Andreev processes~\cite{Andreev} in which electrons tunnel into and out of the superconductor in pairs. 

In the leading order in $E_C/\Delta\ll 1$ the two-particle tunneling amplitudes in \Eq{5} are given by~\cite{HGMS,Garate}
\beq
J_{\alpha\alpha'} 
= \sum_n\frac{\Delta}{\Delta^2 + \varepsilon_n^2}\msk t^\ast_{\alpha n}t^\ast_{\alpha' n}\smsk,
\label{6}
\eeq
and are subject to mesoscopic fluctuations. Provided that the motion of electrons inside the superconductor is chaotic, such fluctuations can be analyzed using the standard random matrix theory-based prescriptions (see, e.g., Refs. \cite{ABG_review,PG_reviews} and references therein). In this approach, the single-particle tunneling amplitudes $t_{\alpha n}$ are statistically independent of each other and of the single-particle energies $\varepsilon_n$. Accordingly, the sum in~\Eq{6} consists of a large (of order \mbox{$\Delta/\delta\gg 1$}) number of statistically independent random contributions. The central limit theorem then suggests that the distribution of $J_{\alpha\alpha'}$ is Gaussian. Using 
\mbox{$\llangle t_{\alpha n}\rrangle = 0$} and 
\mbox{$\llangle t^\past_{\alpha m}t^\past_{\beta n}\rrangle 
= \llangle t^\past_{\alpha m}t^\ast_{\beta n}\rrangle 
= (2\pi)^{-1}\delta v\smsk g_\alpha\delta_{\alpha\beta}\delta_{mn}$}~\cite{ABG_review,PG_reviews}, where the double angular brackets denote averaging over the mesoscopic fluctuations and $g_\alpha$ is the dimensionless (in units of $2e^2\nmsk/h$) conductance of contact $\alpha$, and replacing the summation over $n$ by the integration, we find
\beq
\llangle J_{\alpha\alpha'}\rrangle = J_\alpha  \delta_{\alpha\alpha'},
\quad
J_\alpha = \dfrac{\msk g_\alpha v}{2}
\label{7}
\eeq
and 
\begin{subequations}
\label{8}
\beqa
\llangle J_{\alpha\alpha}J_{\alpha'\alpha'}\rrangle - J_\alpha J_{\alpha^\prime}
&=& \dfrac{1}{2\pi} \dfrac{\delta}{\Delta}\msk(1 + \delta_{\alpha\alpha'}\nmsk)
J_\alpha J_{\alpha'},
\label{8a}\\[4pt]
\llangle J^2_{\alpha\alpha'}\rrangle^\pdag_{\alpha\neq\alpha'}
&=& \dfrac{1}{2\pi} \dfrac{\delta}{\Delta} J_\alpha J_{\alpha'}.
\label{8b}
\eeqa
\end{subequations}
These equations show that both the off-diagonal elements of the $2\snmsk\times\snmsk 2$ matrix $J_{\alpha\alpha'}$ and the fluctuations of the diagonal elements are parametrically suppressed at $\delta/\Delta \ll 1$, and can be neglected. Note that  $\delta/\Delta \ll 1$ is the limit when the BCS description of the superconductor employed in the above derivation is accurate~\cite{vDelft}.

The charge states $|N\pm 1\rangle$ in \Eq{5} are discriminated by electrostatics, see \Eq{3}. For almost all values of the gate potential $N_{\snmsk g}$, the ground state of $H_C$ is nondegenerate. Exceptions are narrow intervals of $N_{\snmsk g}$ around odd integers $N_{\snmsk g}^\ast\nmsk$, where states with \mbox{$N = N_{\snmsk g}^\ast\!\pm 1$} electrons have almost identical electrostatic energies. Accordingly, at \mbox{$T\ll E_C$} and \mbox{$|N_{\snmsk g}^\past\! - N_{\snmsk g}^\ast|\ll 1$} the Hamiltonian can be simplified further by discarding all but the two almost degenerate charge states $|N_{\snmsk g}^\ast\!\pm1\rangle$, which can be viewed as two eigenstates of spin-1/2 operator $S$, \mbox{$|N_{\snmsk g}^\ast +1\rangle \to |\!\!\uparrow\rangle$} and \mbox{$|N_{\snmsk g}^\ast -1\rangle \to |\!\downarrow\rangle$}. Upon performing the particle-hole transformation~\cite{Garate} $c_{\alpha k\downarrow}^\pdag \to c^\dagger_{\alpha, -k,\downarrow}$ and taking into account \Eq{7}, we arrive at the Hamiltonian of the anisotropic two-channel Kondo model~\cite{NB,CZ,SS,Affleck,GNT}
\beq
H = H_0 + B S^z 
+ \sum_\alpha J_\alpha^\pdag \bigl[s_\alpha^+(0)\smsk S^- + s_\alpha^-(0)\smsk S^+\snmsk\bigr],
\label{9}
\eeq
where $s_\alpha^+(0) = c^\dagger_{\alpha\uparrow}(0)\smsk c^\pdag_{\alpha \downarrow}(0)$,
$s_\alpha^-(0) = [s_\alpha^+(0)]^\dagger$, and \mbox{$B = 4E_c\smsk(N_{\snmsk g}^\ast\nmsk - N_{\snmsk g}^\past\!)$}. 
(In writing \Eq{9}, we changed the sign of the exchange term with the help of the unitary transformation $e^{i\pi S^z}\!\nmsk H e^{-i\pi S^z}\!$.) 

Importantly, the exchange constants $J_\alpha$ in \Eq{9} are controlled independently by the conductances of the point contacts [see \Eq{7}], and, therefore, can be easily tuned to be equal. Similarly, the ``magnetic field''~$B$ describes departures from the charge degeneracy and can be tuned to zero by changing the gate potential~$N_{\snmsk g}$. Such remarkable tunability allows one to fully explore various parameter regimes of the two-channel Kondo model~\eq{9}.

At $B = 0$ and $J_L = J_R$ [these equations define a line in the three-dimensional parameter space $(B,J_L,J_R)$] observable quantities exhibit a non-Fermi liquid behavior~\cite{NB,CZ,SS,Affleck,GNT}, whereas anywhere away from this critical line they behave at lowest temperatures as prescribed by the Fermi-liquid theory. On crossing the critical line at $T = 0$, the system undergoes a quantum phase transition between two Fermi-liquid states that are adiabatically connected to each other by going around the critical line. At the transition, observable quantities exhibit singularities. For example, at $J_L = J_R$ the susceptibility \mbox{$d\langle S_z\rangle/dB$}, associated with the correlation function $\langle S_z(t)S_z(0)\rangle$, diverges logarithmically~\cite{SS} at $B \to 0$. 

Our observable of choice, the linear conductance, is given by the Kubo formula~\cite{Mahan}
\beq
\frac{G\,}{\msk G_0} = \lim_{\omega\msk\to\msk 0}\msk
\frac{\pi}{\omega}\nmsk\int_0^\infty\!\!dt \,e^{i\omega t}\snmsk\left\langle [\mathcal I(t),\mathcal I(0)]\right\rangle
\label{10}
\eeq
with $G_0 = 2e^2\nmsk/h$ and with the particle current operator given by
\beq
\mathcal I = \frac{d}{dt}\frac{1}{2}\smsk(\mathcal N_R - \mathcal N_L),
\label{11}
\eeq
where $\mathcal N_\alpha$ is the total number of electrons in the lead~$\alpha$. In terms of the Kondo model \eq{9}, it reads
\beq
\mathcal N_{\alpha\phantom{\uparrow}}^\pdag \!\!\! = N_{\alpha\uparrow}^\pdag - N_{\alpha\downarrow}^\pdag,
\quad
N^\pdag_{\alpha\sigma\phantom{\uparrow}} \!\!\! = \sum^\pdag_{k\sigma} 
c^\dagger_{\alpha k\uparrow} c^\pdag_{\alpha k\uparrow}.
\label{12}
\eeq
With time dependence governed by the Hamiltonian \eq{9}, we find 
\mbox{$\mathcal I = i\smsk \bigl[\snmsk J_L^{\phantom{+}} s_L^+(0) - J_R^{\phantom{+}} s_R^+(0)\snmsk\bigr]\smsk S^- + \text{H.c.}$}.  
Accordingly, the conductance \eq{10} provides direct access to the correlation functions of the type $\langle S^+(t)s_L^-(t)S^-(0)s_R^+(0)\rangle$.

We discuss first the temperature dependence of the conductance when the parameters of the Kondo model~\eq{9} are tuned precisely to the critical line. In other words, we consider exact charge degeneracy $(B = 0)$, and equal conductances of the contacts \mbox{$(J_L = J_R = J = gv/2)$}. 
Writing the rate equations result~\cite{HGMS,HLG} in terms of exchange constants in \Eq{9}, we find $G/G_0 = 2\pi^2(\nu J)^2$ for the conductance in the lowest order in $\nu J\ll 1$ [here \mbox{$\nu = (2\pi v)^{-1}$} is the density of states per length]. 
The Kondo effect can be accounted for in the rate equations formalism~\cite{FM95} by replacing $J$ with its renormalized value reached when the bandwidth $D$ of conduction electrons in \Eq{9} is reduced from its initial value $D\sim E_C$ to~\mbox{$D\sim T$}.  In the scaling limit~\cite{Suppl} \mbox{$T_K\ll D\ll E_C$} we have \mbox{$\nu J(D) = [2\ln(D/T_K)]^{-1}$}, and the conductance assumes the form
\beq
\frac{G\,}{\msk G_0} = \frac{\pi^2}{2{\mkern -1mu}\ln^2(T/\smsk T_K)},
\label{13}
\eeq
where \mbox{$T_K\sim E_C\smsk e^{-\pi^2\nmsk/g}$} is the Kondo temperature~\cite{Matveev91,Suppl}. 

The temperature dependence of the conductance in the strong-coupling regime ($T\ll T_K$) can be found using the technique of Ref.~\cite{EK}, which yields 
\beq
\frac{G\,}{\msk G_0} 
= 1-a\msk\frac{T\msk}{\msk T_K}, \quad a\sim 1.
\label{14}
\eeq 
(The value of $a$ depends on the precise definition of $T_K$). This result can be derived by considering the least-irrelevant perturbation of the Emery-Kivelson Hamiltonian \cite{EK}, which is the same perturbation as the one producing the correct low-temperature asymptote of the specific heat in the two-channel Kondo model~\cite{Sengupta}. A standard perturbative calculation of the conductance then yields~\cite{Mora,Harold} the linear-in-$T$ dependence of $G$.

Alternatively, \Eq{14} can be obtained by mapping~\cite{EA92,YiKane} our problem onto that of a resonant tunneling of a Luttinger liquid with the Luttinger-liquid parameter $K = 1/2$ through a double-barrier structure~\cite{KaneFisher}. Accounting for the least-irrelevant (at $K>1/3$) perturbation identified in Ref.~\onlinecite{EA92}, the correction to the conductance scales as $G(0) - G(T) \propto T^{\smsk 2K}$, in agreement with \Eq{14}. Note that $G(T)$ we found differs from that in the two-channel Kondo device proposed in~Ref.~\cite{OGG} and realized experimentally in~Ref.~\cite{GG}. The difference arises because $G$ in the device of Refs.~\cite{OGG,GG} is proportional to the single-particle \mbox{$t\msk$-matrix}~\cite{we_2CK}, hence~\cite{Affleck,AL} \mbox{$G(0) - G(T)\propto\sqrt{T}$}, whereas in our case $G$ is given by the two-particle correlation function.

According to \Eq{14}, the conductance at zero temperature is exactly half of the conductance of an ideal single-channel interface between a normal conductor and a superconductor~\cite{BTK} $4e^2\nmsk/h$. Such halving of the ideal conductance is one of the manifestations of quantum criticality. This property is reminiscent of the predicted~\cite{FM95,Sela_blockade} and observed~\cite{Pierre2015} behavior of inelastic cotunneling of spin-polarized electrons through a Coulomb-blockaded normal-state island with vanishing single-particle level spacing. Indeed, in this case the zero-temperature conductance at the charge-degeneracy point is \mbox{$e^2\nmsk/2h$}, which again is exactly half of the ideal conductance of a single-channel point contact~$e^2\nmsk/h$.

Finite zero-temperature conductance in our model is the hallmark of the non-Fermi-liquid behavior. Any departure from the critical line restores the Fermi liquid: at finite $B$, $J_L - J_R$, or both, the conductance scales as $G\propto T^2$ at lowest temperatures instead of~\Eq{14}. The origin of this behavior is easy to understand in the limit of large $B\gg T_K$. In this limit, the entire dependence $G(T)$ can be found by perturbation theory. At $T\ll B$ transitions $|\nmsk\!\downarrow\rangle\to|\nmsk\!\uparrow\rangle$ are virtual, and their role reduces to merely generating a residual local exchange interaction between conduction electrons~\cite{NB,Nozieres_FL}. The contribution giving rise to nonzero current reads 
\beq
H_\text{int} = V\smsk\bigl[s^+_R(0) s^-_L(0) + s^-_R(0) s^+_L(0)\snmsk\bigr]. 
\label{15}
\eeq 
In the second order of perturbation theory, the interaction constant in \Eq{15} is given by $V = -\msk J^2\nmsk/B$. With~$J$ here replaced with its renormalized value at \mbox{$D\sim B$}, \Eq{15} is applicable at all $B$ in the range \mbox{$T_K\!\ll B\ll E_C$}. The particle current [see Eqs.~\eq{11} and \eq{12}] evaluated with the Hamiltonian \mbox{$H = H_0 + H_\text{int}$} reads
\mbox{${\mathcal I} = 2iV[(s^+_L(0) s^-_R(0)  - s^+_R(0) s^-_L(0)]$}. The Kubo formula~\eq{10} then yields $G/G_0 = [(2\pi)^4\nmsk/3]\smsk\nu^4 V^2 T^2$, leading to the asymptote 
\beq
\frac{G\,}{\msk G_0} = \frac{\pi^4 \smsk T^2}{3 B^2\snmsk\ln^4(B/\smsk T_K)}
\label{16}
\eeq 
at \mbox{$T\ll B$}. On the other hand, in the opposite limit $T\gg B$ the conductance is still described by \Eq{13}. Hence, the dependence $G(T)$ is nonmonotonic, with a maximum at $T\sim B$.

\begin{figure}
\centerline{\includegraphics[width=\columnwidth]{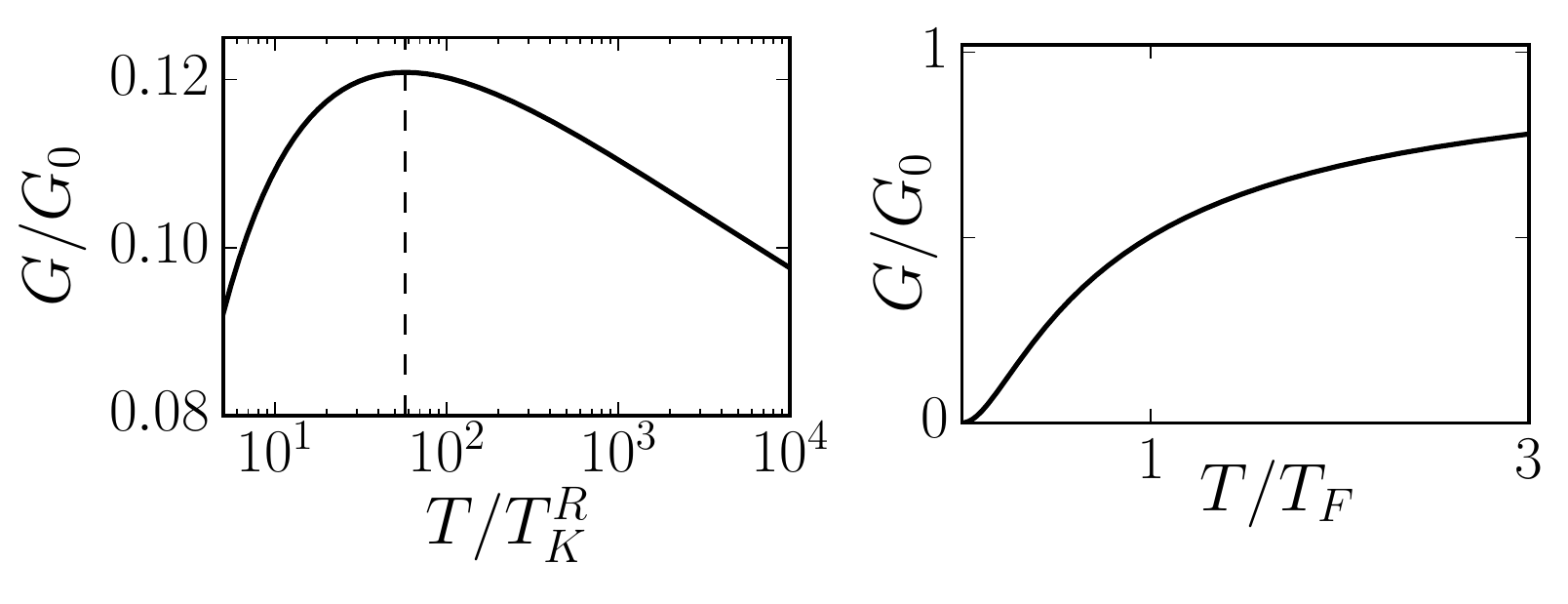}}
\vspace{-3ex}
\caption{
Left panel: Nonmonotonic temperature dependence of the conductance at the Coulomb-blockade peak for asymmetric contacts with $g_L=0.2$ and $g_R=0.3$, corresponding to $T_K^L/T_K^R\sim 10^{-8}$. 
Right panel: Temperature dependence of the conductance in the universal regime $T\sim T_F\ll T_K$, see \Eq{18}. 
}
\label{figure}
\end{figure}

The channel asymmetry also leads to a nonmonotonic temperature dependence of the conductance. If the contact conductances are small but very different, the conductance reaches its maximum in the regime accessible by perturbative renormalization. To be definite, we consider the case when $J_L\ll J_R$. Evaluating the conductance with the help of the rate equations~\cite{HGMS,HLG}, we find $G/G_0=8\pi^2(\nu J_L)^2$. When considering perturbative renormalization of $J_L$~\cite{NB,CZ,Suppl,Matveev91}, it is important to take into account, in addition to the usual second-order term $J_L^2$, the dominant next-order contribution. This contribution is proportional to $J_LJ_R^2$ and is negative, leading to a nonmonotonic dependence $J_L(D)$~\cite{Suppl}. In the scaling limit, it is convenient to express the results in terms of $T_K^{\alpha}\sim E_C\exp(-\smsk\pi^2\nmsk/g_{\alpha})$ ($T_K^\alpha$ is the Kondo temperature in the limit when the conductance of contact~$\alpha$ is finite, whereas the second contact is completely shut off). The conductance reaches its maximum
\beq
\left(\frac{G\,}{\msk G_0}\snmsk\right)_{\nmsk\text{max}} \! 
= \frac{\smsk 2\pi^2}{\ln^2\bigl(T_K^R/\msk T_K^L\bigr)}
\label{17}
\eeq
at $T\sim T_K^R\smsk\exp{\sqrt{\smsk\ln\smsk\bigl(T_K^R/\msk T_K^L\bigr)}}$~\cite{Suppl}, which belongs to the perturbative domain $(T\gg T_K^R\sim T_K^\pdag)$ provided that $|g_R-g_L|\gg g_R^2$. The dependence on temperature near the maximum is weak, see the left panel in \Fig{figure}, and $G(T)$ crosses over from \Eq{17} to the Fermi-liquid low-temperature asymptote $G\propto T^2$ at $T\sim T_K^R$.

In the opposite limit of small deviations from the critical line in the parameter space $(B,J_L,J_R)$, the system upon lowering the temperature first enters the strong-coupling non-Fermi-liquid regime [see \Eq{14}], and then crosses over at $T\sim T_F\ll T_K$ to the limiting Fermi-liquid behavior. The crossover is described by~\cite{FM95,Sela_blockade} 
\beq
\frac{G\,}{\msk G_0} 
= f\left(\frac{\pi T\,}{\msk T_F}\right), 
\quad
f(x) = 1 - \frac{1}{2x}\msk\Psi\left(\frac{1 + x}{2x}\right),
\label{18}
\eeq
where $\Psi(z) = d^{\msk 2}\nmsk\ln\Gamma(z)/dz^2$ is the trigamma function. The universal function $f(x)$ interpolates between \mbox{$f(x) = 1 - \pi^2\nmsk/4 x$} at $x\gg 1$ and  \mbox{$f(x) = x^2$} at $x\ll 1$. The latter limit corresponds to the Fermi-liquid regime. The dependence $G(T)$ given by \Eq{18} is plotted in the right panel in \Fig{figure}.

The characteristic crossover scale $T_F$ in \Eq{18} is set by the distance of the system parameters to the critical line. This scale can be estimated by scaling analysis. Near the critical line, both the magnetic field (i.e., distance to the charge degeneracy point) and the channel asymmetry are relevant perturbations with scaling dimension $1/2$~\cite{AL91,2CK_asymmetry,Zarand,Sela}. Therefore, as the bandwidth is lowered, the corresponding dimensionless coupling constants 
grow at $D\lesssim T_K$ as $(T_K/D)^{1/2}$, becoming of order unity at $D\sim T_F$. Taking into account that $B$ at $D\sim T_K$ is of order of its bare value~\cite{Suppl}, we obtain~\cite{CZ,Zarand} \mbox{$T_F \sim B^2\nmsk/T_K \sim (E_C^{\msk 2}\smsk/\smsk T_K)(N_{\snmsk g}^\past\! - N_g^\ast\nmsk)^2$} for channel-symmetric setup~\cite{T_F}. Accordingly, the width of the Coulomb blockade peak in the dependence $G(N_{\snmsk g}\snmsk)$ scales as $\sqrt{T}$ with temperature. The above estimate of~$T_F$ and \Eq{18} are applicable as long as $T_F\ll T_K$, i.e., close to the charge degeneracy point. Further away from this point, the conductance is described by Eqs.~\eq{13} and \eq{16} at $T\gg B$ and $T\ll B$, respectively. 

Interestingly, \Eq{18} also describes conductance in a device with almost open contacts, i.e., in the limit when $1 - g_\alpha\ll 1$ and the tunneling Hamiltonian description of the contacts [see \Eq{4}] is inapplicable. In fact, it was originally derived~\cite{FM95} in this limit in the context of the closely related problem of inelastic cotunneling. For almost open contacts the crossover scale $T_F$ also scales as $(N_{\snmsk g}^\past\! - N_g^\ast\nmsk)^2$ in the vicinity of the charge degeneracy point~\cite{FM95}; hence, the width of the Coulomb blockade peak is again proportional to~$\sqrt{T}$. However, in this limit the number of electrons in the Coulomb-blockaded region is not quantized. Strong charge fluctuations render the reduction to the Kondo model [cf.~\Eq{9}] impossible. As a result, the temperature dependence of the conductance is characterized by only two energy scales, $E_C$ and~$T_F$~\cite{FM95}.
In the symmetric case, despite the absence of the intermediate scale $T_K$, the linear-in-$T$ correction to the conductance at the degeneracy point remains valid, except that $T_K$ is replaced by $E_C$ in Eq.~\eq{14}.

In conclusion, conduction through a Coulomb-blockaded mesoscopic $s$-wave superconductor
is facilitated by Andreev processes. In the vicinity of the charge degeneracy points 
these processes can be mapped onto exchange terms in the effective two-channel Kondo model. Unlike in the case of inelastic cotunneling through a normal-state island~\cite{Matveev91,FM95,Sela_blockade,Pierre2015}, the mapping does not rely on the smallness of the single-particle level spacing in the Coulomb-blockaded region in comparison with temperature. The critical two-channel-Kondo regime corresponds to the limit when conductances of the point contacts connecting the superconductor to the normal-state leads are equal. In such symmetric setup conductance at the Coulomb blockade peak increases with the decrease of temperature, reaching $2e^2\nmsk/h$ at zero temperature, whereas the width of the peak decreases as~$\sqrt{T}$. 

Our theory is valid provided that the induced superconducting gap $\Delta$ is large compared with both the charging energy $E_C$ and the single-particle level spacing $\delta$. These parameters are set by the device geometry and properties of the materials used. Experiments~\cite{Albrecht2016} on the already existing $1.5\,\mu\smsk\text{m}$-long aluminum-coated InAs wires yielded $\Delta\approx 0.2\,\smsk m\smsk\text{eV}$ and \mbox{$E_C\approx 20\,\mu\smsk\text{eV}$}. Estimating $\delta$ with the help of the results of Ref.~\cite{HLG}, we find $\delta\sim 3\,\mu\smsk\text{eV}$. Accordingly, parameters of these wires fall well within the desired range. Estimated values of these parameters for the prospective devices~\cite{Pierre_private} of the type studied in Ref.~\cite{Pierre2015} with normal metal NiGeAu replaced by superconducting In read \mbox{$\Delta\sim 1\,m\smsk\text{eV}$, $E_C\sim 20\,\mu\smsk\text{eV}$, and $\delta\sim 10^{-5}\,\mu\smsk\text{eV}$,} thus promising a much larger value of the ratio $\Delta/\delta$. Unlike $\Delta$, $E_C$, and~$\delta$, the Kondo temperature $T_K$ and the crossover scale~$T_F$ are tunable by varying conductances of the contacts and the gate potential. The tunability makes it possible to explore experimentally all the regimes discussed above and crossovers between them on a single device.

\begin{acknowledgments}
We thank Harold Baranger, Christophe Mora, and Eran Sela for pointing out the correct temperature dependence of Eq.~\eqref{14}, which corrects a previous version of the manuscript, and for helping us understand its origin.
We are grateful to Fabrizio Nichele and Frederic Pierre for discussions and correspondence.
This work is supported by ONR Grant Q00704 (BvH) and by DOE contract
DEFG02-08ER46482 (LG).
\end{acknowledgments}
\onecolumngrid

\onecolumngrid
\clearpage
\begin{center}
{
{\large \textbf{Quantum Criticality in Resonant Andreev Conduction}}
\\[10pt]
\textbf{Supplemental Material}
}

\thispagestyle{empty}

\vspace{0.15in}

M. Pustilnik,$^1$ B. van Heck,$^2$ R. M. Lutchyn,$^3$ and L. I. Glazman$^2$

\vspace{0.075in}
\small\textit{
$^\textit{1}$School of Physics, Georgia Institute of Technology, Atlanta, Georgia 30332, USA
\\[.5pt]
$^\textit{2}$Department of Physics, Yale University, New Haven, Connecticut 06520, USA 
\\[.5pt]
$^\textit{3}$Station Q, Microsoft Research, Santa Barbara, California 93106-6105, USA
}
\end{center}

\vspace{-2ex}

\numberwithin{equation}{section}

\vspace{4ex}
\centerline{\color{blue}\textbf{
1. Two-channel Kondo model in the weak coupling regime
}}
\vspace{3ex}
\twocolumngrid
\setcounter{section}{1}
\setcounter{equation}{0}

We consider the two-channel Kondo model  
\beqa
H\!&=&\!\sum_{\alpha \sigma}\int\!dx\,\colon c^\dagger_{\alpha\sigma}(x)\smsk(-iv\partial_x)\msk c^\pdag_{\alpha\sigma}(x)\msk\colon 
\,+\msk BS^z
\label{1.1}\\[2pt]
&&\,
+ \sum_\alpha\,\Bigl\{J_{z\alpha}^\pdag\smsk s_\alpha^z(0)\smsk S^z 
+ J_{\perp\alpha}\msk\bigl[s_\alpha^+(0)\smsk S^- + s_\alpha^-(0) \smsk S^+\bigr]\Bigr\},
\nn
\eeqa
where \mbox{$c_{\alpha\sigma}(x) = L^{-1/2}\nmsk\sum_k\snmsk e^{ikx} c_{\alpha k}$}, the colons denote the normal ordering, and
\begin{subequations}
\label{1.2}
\beqa
s_\alpha^z(0) \!&=&\! \frac{1}{2}\msk
\bigl[\colon c^\dagger_{\alpha\uparrow}(0)\msk c^\pdag_{\alpha\uparrow}(0)\msk\colon
\;-\; \colon c^\dagger_{\alpha\downarrow}(0)\msk c^\pdag_{\alpha\downarrow}(0)\msk\colon
\nmsk\bigr]\smsk,
\qquad
\label{1.2a}\\[2pt]
s_\alpha^+(0) \!&=&\! \bigl[s_\alpha^-(0)\bigr]^\dagger =  c^\dagger_{\alpha\uparrow}(0)\msk c^\pdag_{\alpha\downarrow}(0)\smsk.
\label{1.2b}
\eeqa
\end{subequations}
The exchange amplitudes corresponding to the initial bandwidth \mbox{$D_0\sim E_C$} are given by
\beq
J_{z\alpha} = 0, 
\quad 
J_{\perp\alpha} = \frac{g_\alpha v}{2}\msk, 
\label{1.3}
\eeq
see 
Eq.~(7) in the paper. 
Upon reduction of the bandwidth $D$, the dimensionless exchange amplitudes
\beq
I_{z\alpha} = \frac{\smsk J_{z\alpha}}{2\pi v}, 
\quad
I_{\perp\alpha} = \frac{J_{\perp\alpha}}{\pi v} 
\label{1.4}
\eeq
evolve according to the weak-coupling renormalization group equations~\cite{sNB,sCZ,sMatveev91} 
\begin{subequations}
\label{1.5}
\beqa
\frac{d}{d\zeta} I_{z\alpha} &=& I_{\perp\alpha}^2 - \frac{1}{2}\smsk I_{z\alpha}\sum_\beta I_{\perp\beta}^2
+\ldots,
\label{1.5a} \\
\frac{d}{d\zeta} I_{\perp\alpha}&=& I_{z\alpha} I_{\perp\alpha} 
- \frac{1}{4}\smsk I_{\perp\alpha}\sum_\beta \bigl(I_{\perp\beta}^2 + I_{z\beta}^2\bigr)+\ldots,
\quad
\label{1.5b}
\eeqa
\end{subequations}
where $\zeta = \ln\smsk(D_0/D)$. Neglecting the cubic terms in the right-hand sides of Eqs.~\eq{1.5}, we obtain~\cite{sMatveev91}
\begin{subequations}
\label{1.6}
\beqa
I_{\perp\alpha}(\zeta) &=& \frac{I_{\perp\alpha}(0)}{\sin[I_{\perp\alpha}(0)(\zeta_\alpha - \zeta)]}\msk,
\label{1.6a} \\[6pt]
I_{z\alpha}(\zeta) &=& I_{\perp\alpha}(0)\cot[I_{\perp\alpha}(0)(\zeta_\alpha - \zeta)]
\label{1.6b}
\eeqa
\end{subequations}
with
\beq
\zeta_\alpha = \frac{\pi}{2I_{\perp\alpha}(0)}\msk.
\label{1.7}
\eeq
The equation $\ln(D_0/\smsk T^{\smsk\alpha}_K) = \zeta_\alpha$ gives the estimate of the Kondo temperature in the channel $\alpha$ when the other channel is completely shut off,
\beq
T_K^{\smsk\alpha} \sim D_0\smsk e^{-\smsk\zeta_\alpha} 
\sim E_C\smsk e^{-\smsk\pi^2\nmsk/g_\alpha}.
\label{1.8}
\eeq
[Corrections to $T_K^\alpha$ come from the cubic and higher-order terms in the right-hand sides of Eqs.~\eq{1.5} neglected in the derivation of Eqs.~\eq{1.6}.] 

In the scaling limit defined by
\beq
1\ll \zeta_\alpha - \zeta \ll \zeta_\alpha
\label{1.9}
\eeq
Eqs.~\eq{1.6} simplify to
\beq
I_{\perp\alpha}(\zeta) = I_{z\alpha}(\zeta) = I_\alpha(\zeta) 
= \frac{1}{\zeta_\alpha - \zeta}
= \frac{1}{\ln(D/T_K^{\smsk\alpha})}\smsk, 
\label{1.10}
\eeq
indicating a restoration of $SU(2)$ symmetry. In the channel-symmetric case we have 
$\zeta_\alpha = \zeta_K$ and $T_K^\alpha = T_K^\pdag$. In terms of $D$ and $T_K$, the condition \eq{1.9} then reads
\beq
T_K\ll D\ll E_C,
\label{1.11}
\eeq 
and \Eq{1.10} assumes the form
\beq
I_\alpha (D) = \frac{1}{\ln(D/T_K)}\smsk.
\label{1.12}
\eeq
With the identification $\nu J_\alpha = I_\alpha/\smsk 2$ [see \Eq{1.4}], this expression is used in 
Eqs.~(13) and (16) in the paper.

Renormalization of the magnetic field $B$ in \Eq{1.1} is governed by the equation
\beq
\frac{d}{d\zeta} B = -\,\frac{B}{2}\sum_\alpha I_{\perp\alpha}^2(\zeta)+\ldots\msk.
\label{1.13}
\eeq
Taking into account  Eqs.~\eq{1.6}, we find
\beq
\ln \frac{B(0)}{B(\zeta)} = \frac{1}{2}\sum_\alpha I_{z\alpha}(\zeta)\smsk. 
\label{1.14}
\eeq
For symmetric channels, \Eq{1.14} reduces in the scaling limit to
\beq
\ln \frac{B(0)}{B(\zeta)} = \frac{1}{\zeta_K - \zeta}\msk.
\label{1.15}
\eeq
The right-hand side of \Eq{1.15} is small at all~\mbox{$\zeta\ll\zeta_K$}, becoming of order unity only at \mbox{$\zeta_K - \zeta \sim 1$}, when Eqs.~\eq{1.5} and \eq{1.13} cease to be applicable. Therefore, the renormalization of $B$ throughout the weak coupling regime $\zeta_K - \zeta\gg 1$ can be ignored in the first approximation. This lack of renormalization is taken into account in writing 
Eq.~(16) in the paper.

If the asymmetry between the channels is strong, e.g., $J_L\ll J_R$ or, equivalently, $\zeta_L\gg\zeta_R$, the exchange amplitudes grow with $\zeta$ according to Eqs.~\eq{1.6} until $\zeta$ reaches the value $\zeta_\ast$ at which the cubic terms in Eqs.~\eq{1.5} for the weaker coupled channel become compatible with the quadratic ones. If $\zeta_\ast$ belongs to the scaling limit [see \Eq{1.9}], this leads to the equation \mbox{$I_L^\pprime(\zeta_\ast\nmsk) = I_R^{\smsk 2}(\zeta_\ast\nmsk)$}, where $I_\alpha(\zeta)$ are given by \Eq{1.10} with $D = T_\ast$. Taking into account that $\zeta_L - \zeta_\ast\gg\zeta_R-\zeta_\ast\gg 1$, we obtain 

\vspace{4ex}
\begin{subequations}
\label{1.16}
\beqa
&\nn\\[-8ex]
&\ds I_L(\zeta_\ast) = \frac{1}{\,\ln\bigl(T_K^R/\msk T_K^L\bigr)}\smsk,~
\label{1.16a}\\[3pt]
&\ds\ln\smsk\bigl(T_\ast/\msk T_K^R\bigr)\sim \bigl[\ln\smsk\bigl(T_K^R/\msk T_K^L\bigr)\bigr]^{1/2}.
\label{1.16b}
\eeqa
\end{subequations}
With further reduction of the bandwidth, at \mbox{$\zeta_\ast\ll\zeta\ll\zeta_R\sim\zeta_K$}, the smaller exchange amplitude~$I_L$ evolves according to the equation
\beq
\frac{d}{d\zeta} I_L = -\msk \frac{1}{2}I_L^\pprime\nmsk I_R^{\smsk 2}
\label{1.17}
\eeq
which describes a downward renormalization of $I_L(\zeta)$ similar to that of $B(\zeta)$. Accordingly, $I_L(\zeta)$ is nonmonotonic, with \mbox{$\max\{I_L(\zeta)\snmsk\} = I_L(\zeta_\ast)$}. The estimates~\eq{1.16} are used in 
Eq.~(17) in the paper.

\onecolumngrid
\vspace{-1ex}


\begin{thebibliography}{99}
\bibitem{footnote} 
Note that the recently developed field-theoretical~\cite{BC} and numerical~\cite{PZL} models predict that more sophisticated multi-terminal devices carrying multiple Majorana modes have highly nontrivial properties, drastically different from two-terminal devices.
\bibitem{T_F}
At the charge degeneracy point the crossover scale in \Eq{18} is controlled by the channel asymmetry and can be estimated as 
\mbox{$T_F \sim [(g_L - g_R)^2\nmsk/(g_L + g_R)^4]\msk T_K$}~\cite{we_2CK}.
\end{thebibliography}
\end{document}